# Ab initio study of the alloying effect of transition metals on structure, stability and ductility of CrN

Liangcai Zhou,[1, a)] David Holec,[2] and Paul. H. Mayrhofer[1, 3]

[1]*Institute of Materials Science and Technology, Vienna University of Technology, A-1040 Vienna, Austria*

[2]*Department of Physical Metallurgy and Materials Testing, Montanuniversität Leoben, A-8700 Leoben, Austria*

[3]*Christian Doppler Laboratory for Application Oriented Coating Development at the Institute of Materials Science and Technology, Vienna University of Technology, A-1040 Vienna, Austria*

Abstract:

The alloying effect on the lattice parameters, isostructural mixing enthalpies and ductility of the ternary nitride systems $Cr_{1-x}TM_xN$ (TM=Sc, Y; Ti, Zr, Hf; V, Nb, Ta; Mo, W) in the cubic B1 structure has been investigated using first-principles calculations. Maximum mixing enthalpy due to large lattice mismatch in $Cr_{1-x}Y_xN$ solid solution shows a strong preference for phase separation, while $Cr_{1-x}Ta_xN$ exhibits a negative mixing enthalpy in the whole compositional range with respect to cubic B1 structured CrN and TaN, thus being unlikely to decompose spinodally. The near-to-zero mixing enthalpies of $Cr_{1-x}Sc_xN$ and $Cr_{1-x}V_xN$ are ascribed to the mutually counteracted electronic and lattice mismatch effects. Additions of small amounts of V, Nb, Ta, Mo or W into CrN coatings increase its ductility.

[a)]Author to whom correspondence should be addressed;

E-mail address: liangcai.zhou@tuwien.ac.at

## 1. Introduction

The application of transition metal nitrides hard coatings to protect cutting tools and mechanical components has become a common practice in the manufacturing industry in the last two decades [1]. Conventional binary hard coatings such as TiN and CrN are often used to enhance the performance of cutting tools [2]. Due to the superior oxidation resistance of CrN when compared` with TiN, CrN is one of the most important transition metal nitrides, and highly valued as a protective and anti-wear coating [3].

In order to further tune the structural, mechanical, and tribological properties, i.e., incorporation of other elements into the CrN based coating, alloying proves to be an effective concept [4-7]. Recently, various $Cr_{1-x}X_xN$ (X=Al, Ti, Ta, Zr, Mo, W and V) ternary coatings have received lots of attention due to their excellent properties [4-6, 8-16]. One of the most successful examples is the addition of Al to improve mechanical properties as well as oxidation resistance of CrN [6]. It is therefore not surprising that this system has been heavily studied both experimentally and theoretically [17-25]. Addition of V, Mo or W into CrN improves its tribological properties and toughness [11-13, 16], while thermal stability and magnetic properties of CrN can be adjusted by alloying it with TiN, hence forming a $Cr_{1-x}Ti_xN$ solid solution [10]. $Cr_{1-x}Ta_xN$ has enhanced simultaneously mechanical and oxidation properties with respect to CrN [5]. No experimental data are available at present for $Cr_{1-x}TM_xN$ (TM=Sc, Y and Hf), and only few studies on CrNbN have been reported in the literature [4, 26].

Apart from the experimental reports for $Cr_{1-x}TM_xN$ systems, the theoretical calculations are limited to only $Cr_{1-x}Ti_xN$ [27]. Recently, plenty of examples have illustrated that a combination of theoretical studies with experimental work represent a successful approach to gain deeper understanding of the material behavior, which can be used in e.g. designing coatings with application tailored properties [6, 18, 19, 28]. For this reason we have carried out a systematic theoretical study of the alloying effect of transition metals (TM=Sc, Y; Ti, Zr, Hf; V, Nb, Ta; Mo, W) on CrN.

## 2. Calculation methods

Density Functional Theory (DFT) based calculations are performed using the Vienna Ab initio Simulation Package (VASP) [29, 30]. The ion-electron interactions are described by the projector augmented wave method (PAW) [31] and the generalized gradient approximation (GGA) is employed for the exchange-correlation effects, as parameterized by Perdew–Burke–Ernzerhof (PBE) [32].

In order to simulate the chemical disorder between Cr and TM atoms on the metal sublattice of the cubic B1 structure (NaCl prototype, space group $Fm\bar{3}m$), and also the paramagnetic state induced by Cr atoms, we use the special quasi-random structures (SQS) [33] approach as implemented in our recent study of $Cr_{1-x}Al_xN$ system [21]. 3×3×2 supercells (36 atoms) described in detail in Ref. [21] are used for the $Cr_{1-x}TM_xN$. Mixing of $Cr^{\uparrow}$, $Cr^{\downarrow}$ and TM atoms takes place on one sublattice while the other sublattice is fully occupied with N atoms. The short range order parameters (SROs) are optimized for pairs at least up to the fifth order. The alloying effect on the ductility has been assessed by evaluating elastic properties of $Cr_{0.89}TM_{0.11}N$ using stress-strain method [21]. All the calculations are performed with plane wave cutoff energy of 500 eV together with 6×6×9 Monkhorst-Pack $k$-point meshes, which guarantee the total energy accuracy in the order of meV per atom.

## 3. Results and discussion

## 3.1. Structural properties

The supercell volume and shape, as well as internal atomic positions are optimized with respect to total energy in order to obtain equilibrium properties of cubic B1 $Cr_{1-x}TM_xN$ solid solutions using the Birch–Murnaghan equation of state [34]. The resulting lattice parameters for $Cr_{1-x}TM_xN$ phases are shown in Fig. 1. Please note that Fig. 1 is divided into a, b, c, and d, according to the TM groups, IIIB, IVB, VB, and VIB, respectively. The calculated values are fitted with a quadratic polynomial:

$$a(x)= xa_{TMN}+(1-x)a_{CrN} +bx(1-x) \qquad (1)$$

where $b$ is a bowing parameter describing the deviation from linear, Vegard's-like behavior [35]. The bowing parameter, $b$, for each $Cr_{1-x}TM_xN$ system together with the optimized lattice constants, $a_0$, their experimental values, $a_{exp}$, for comparison [36, 37], and energies of formation, $E_f$, for TMN compounds are listed in Table 1. The calculated lattice parameters result in slightly larger values than the experimental ones (with the exception of VN), which is expected behavior for DFT-GGA calculation. Good agreement with previous theoretical calculations [17, 38] is also obtained. One should notice that the experimental lattice constants of cubic B1 MoN and WN denote the lattice constants of their substoichiometric N-deficient configurations, as the stoichiometric configurations are mechanically unstable [39, 40]. Energies of formation in Table 1 express the total energy difference between the binary nitride and the corresponding elements in their ground-state configurations (crystalline TM or $N_2$ molecule). The thus obtained trends are the same as reported before by Rovere *et al.* and Holec *et al.* [17, 38]. $E_f$ becomes less negative as the valence electron concentration (VEC) of TM increases. The positive energies of formation of MoN and WN further illustrate their instability.

The bowing parameters, $b$, for the $Cr_{1-x}TM_xN$ (TM=Y, Zr, Hf, Nb and Ta) solid solutions have positive values, which denotes a positive deviation (i.e., to large values) from Vegard's linear interpolation. A positive deviation can be qualitatively rationalized in these cases where the lattice spacings of the constituents are very different. Compressing the larger compound is energetically more costly than to expand the smaller one, due to the anharmonicity of the binding energy curve [41]. For other $Cr_{1-x}TM_xN$ (TM=Sc, Ti, V, Mo and W) solid solutions, the bowing parameters show small negative values. This negative deviation from Vegard's linear interpolation is caused by the stronger interatomic bonds connected with an ordering tendency, hence decreasing the lattice parameter for the intermediate compositions. Especially, the lattice parameters of $Cr_{1-x}Sc_xN$, $Cr_{1-x}Ti_xN$ and $Cr_{1-x}V_xN$ show almost a linear behavior, which can be ascribed to the same $sp^3d^2$ hybridization in the $3d$ group TMNs, consisting of two $3d$ and one $s$ electrons form TM site and three $2p$ electrons donated by nitrogen atom. Comparing the calculated lattice parameters with the experimental data yields good agreement for $Cr_{1-x}TM_xN$ (TM=Zr, Ti, Ta and Nb). In the case of $Cr_{1-x}Zr_xN$, the experimental data of Kim *et al*. [8] confirm our predictions for positive bowing of the $Cr_{1-x}Zr_xN$ lattice parameter, although the deviation from the Vegard's line predicted here seems to be smaller than that observed experimentally. This disagreement could be caused by residual stresses present in the thin films. The calculated lattice parameters of $Cr_{1-x}Ti_xN$, $Cr_{1-x}Nb_xN$ and $Cr_{1-x}Ta_xN$ agree well with the experimentally observed values [4, 5, 10], aside from the fact that the calculated values always show slightly larger values than the experimental ones. In addition to the overestimation of lattice constants with respect to experiment when GGA is used, substoichiometry of nitrogen reported in the experiment might be an important factor. Panel (d) in Fig. 1 shows the optimized lattice parameters of $Cr_{1-x}Mo_xN$ and $Cr_{1-x}W_xN$ solid solutions in comparison with experimental data [4, 12-15]. Because of the mechanical instability of cubic B1 MoN and WN, only the calculated results for the TM content $x$ between 0 to 0.56 are presented. It follows that the calculated lattice parameters of $Cr_{1-x}Mo_xN$ and $Cr_{1-x}W_xN$ solid solutions are very close to each other. The lattice parameters near TM-rich side agree well with experimental measurements, but the difference between theoretical and experimental results increases as TM content increases. This is more prominent for $Cr_{1-x}Mo_xN$. Such behavior can be explained by large substoichiometry of nitrogen in $Cr_{1-x}Mo_xN$ and $Cr_{1-x}W_xN$ solid solutions as it has been reported [4]. Finally, no experimental data are available for the lattice parameters of ternary $Cr_{1-x}TM_xN$ (TM=Sc, Y, Hf and V) solid solutions, which we report here for the first time. The knowledge of lattice parameters of nitrides as given in Table 1 and Fig. 1, provides useful information for designing coatings and for interpretation of experimental results.

3.2. Phase stability

The isostructural mixing enthalpy, $H_{mix}$, as a function of TMN content in each cubic B1 $Cr_{1-x}TM_xN$ solid solution is calculated as:

$$H_{mix} = E(Cr_{1-x}TM_xN) - x\,E(TMN) - (1-x)\,E(CrN) \tag{2}$$

where $E$(XN) and $E(Cr_{1-x}TM_xN)$ are the total energy of binary cubic XN and ternary $Cr_{1-x}TM_xN$, respectively. Figure. 2 summaries the calculated mixing enthalpies of the $Cr_{1-x}TM_xN$ solid solutions. Please note that Fig. 2 is divided into a, b, c, and d, according to the TM groups, IIIB, IVB, VB, and VIB, respectively. Compared with other $Cr_{1-x}TM_xN$ solid solutions, $Cr_{1-x}Y_xN$ displays much larger positive mixing enthalpy with a maximum value of about 0.16 eV/atom at $x \approx 0.55$, indicating a thermodynamic driving force for phase separation into its binary constituents. Recently, Rovere *et al.* [18] reported a considerably larger maximum mixing enthalpy of 0.24 eV/atom, which is likely to be a supercell size effect already which has been pointed out by Žukauskaitė *et al.* [42]. The large positive values of $H_{\text{mix}}$ are the consequence of the large lattice mismatch of YN and CrN, as shown in Fig. 1a. Hence, $Cr_{1-x}Y_xN$ will not tend to mix under realistic equilibrium conditions. On the other hand, the mixing enthalpies of $Cr_{1-x}Sc_xN$ exhibit values very close to zero. The prominently different behavior for the mixing enthalpies between isoelectronic $Cr_{1-x}Y_xN$ and $Cr_{1-x}Sc_xN$ can be traced down to an interplay between lattice mismatch and the electronic effect. The bonding in the TMN can be divided into the hybridized $sp^3d^2$, N-*p*-TM-*d* interaction and metal-metal *d-d* bonding [21, 43]. The stability is related to the extent of the populated nonbonding or antibonding states as reflected by the density of states (DOS) at Fermi level ($E_F$) [17, 19]. High DOS at $E_F$ suggests an increasing destabilizing effect due to the population of energetically unfavorable states. In the case of ScN [44] and YN [42] where small band gap can be found at $E_F$, while negligible DOS or small gap at $E_F$ depending on the exchange-correlation functional in CrN [27] is ascribed to the magnetic spin polarization of spin-up and spin-down states. Hence, the addition of Y [18] and Sc into CrN has a stabilizing effect for the CrN due to depletion of non-bonding or antibonding states. This electronic stabilizing effect is compensated or overcompensated by the lattice strain in the $Cr_{1-x}Sc_xN$ or $Cr_{1-x}Y_xN$, respectively, resulting in near-to-zero mixing enthalpies of $Cr_{1-x}Sc_xN$, and large positive mixing enthalpies in the $Cr_{1-x}Y_xN$ solid solution.

The mixing enthalpy of $Cr_{1-x}Zr_xN$ is much larger than that of $Cr_{1-x}Ti_xN$ and $Cr_{1-x}Hf_xN$ in addition to a strongly asymmetric shape, see Fig. 2b. The maximum values are close to 0.08, 0.03 and 0.02 eV/atom for $Cr_{1-x}Zr_xN$, $Cr_{1-x}Hf_xN$ and $Cr_{1-x}Ti_xN$, respectively, at a TM content around 0.6. It becomes obvious that the larger lattice mismatch between ZrN and CrN as compared with CrN-TiN and CrN-HfN is responsible for the large positive mixing enthalpies of $Cr_{1-x}Zr_xN$. In contrast with the electronic stabilizing effect of Sc and Y, the addition of IV, V and VI B group TM elements into CrN causes destabilization effect due to the surplus $d$ electrons localized at TM sites, which results in increasing of DOS at $E_F$ and consequently large electronic driving force for decomposition at high TM content. $Cr_{1-x}V_xN$ displays similar magnitude of mixing enthalpies close to zero as one can see in Fig. 2c, which is ascribed to the electronic destabilizing effect counteracted by the much smaller lattice mismatch between VN and CrN. $Cr_{1-x}Ta_xN$ is the only system in the present study displaying negative $H_{mix}$ in the full compositional range, indicating TaN and CrN binaries are soluble. However, one should be careful about interpreting the absolute values of $H_{mix}$. The cubic B1 structures are only metastable states of NbN and TaN. When the thermodynamical ground state of TaN i.e., the hexagonal $B_h$ structure (TaN prototype, space group $P6/mmm$), is taken as a reference state, positive values of $H_{mix}$ are obtained also for $Cr_{1-x}Ta_xN$. This indicates that the system is actually thermodynamically unstable, but the decomposition is unlikely to proceed via spinodal process (i.e., decomposition into cubic Cr- and Ta-rich phases).

In the case of $Cr_{1-x}Mo_xN$ and $Cr_{1-x}W_xN$ solid solutions, only results in a limited compositional range are presented here. In Fig. 3d, two different sets of reference states for calculating the mixing enthalpy of $Cr_{1-x}Mo_xN$ and $Cr_{1-x}W_xN$ are chosen. The solid lines denote the mixing enthalpy with cubic CrN and TMN as the reference states, while the dotted lines denote mixing enthalpy with respect to CrN and $TM_2N+N_2$. From Fig. 3d, one can see that the mixing enthalpies of $Cr_{1-x}TM_xN$ calculated with respect to cubic B1 CrN and TMN show more negative values as TM content increases, which suggests increasing stability of these ternary alloys. However, it has been experimentally demonstrated that the deposited $Cr_{1-x}Mo_xN$ and $Cr_{1-x}W_xN$ coatings are mixtures of CrN and $TM_2N$ (TM=Mo and W) [9]. In fact, substoichiometry of nitrogen is always experimentally observed in $Cr_{1-x}Mo_xN$ and $Cr_{1-x}W_xN$ solid solutions, and also the stable cubic $Mo_2N$ and $W_2N$ compounds can be detected in the deposited coatings. We therefore suggest that using $CrN+TM_2N+N_2$ as reference states is more meaningful. The mixing enthalpies of $Cr_{1-x}Mo_xN$ and $Cr_{1-x}W_xN$ with respect to $CrN+TM_2N+N_2$ have positive values (see Fig. 2d), which denotes the possibility for isostructural decomposition, and is consistent with the experimental findings [9]. More detailed results on $Cr_{1-x}TM_xN_{1-y}$ will be presented elsewhere.

3.3 Alloying influence on mechanical behavior

In order for evaluating the alloying effect on the ductility, the bulk-to-shear modulus ratio, $B/G$, and also Cauchy pressure, $C_{12}$-$C_{44}$, as a function of valence electron concentration (VEC) are plotted in Fig. 3. VEC is calculated as an average value of valence electrons per formula unit. According to Pugh *et al*. [45], the higher or lower the $B/G$ ratio is, the more ductile or brittle the material is, respectively. The critical value which separates ductile and brittle materials is approximately 1.75. Figure. 3a demonstrates that the $B/G$ ratio of the ternary $Cr_{0.89}TM_{0.11}N$ solid solutions increases as the VEC increases, which has the same trend with another similar coating system TiTMN. The alloying effect is evaluated through comparing the $B/G$ ratio of $Cr_{0.89}TM_{0.11}N$ with that of CrN. It is clear that adding small amounts of V, Nb, Ta, Mo and W into CrN coatings will increase $B/G$ ratio above 1.75, which denotes the improved ductility. The Cauchy pressure, $C_{12}$-$C_{44}$, can be used to characterize the bonding type [46]. Negative Cauchy pressure corresponds to more directional, while positive values indicate predominant metallic bonding. As the VEC is increased, Cauchy pressure values are increased, which is obviously showing the same trend with $B/G$ ratio for the $Cr_{0.89}TM_{0.11}N$ and suggests the gradual changes in the bonding from directional bonding towards metallic bonding type. A comparison of CrN with $Cr_{0.89}TM_{0.11}N$ yields that alloying V, Nb, Ta, Mo, and W increases the metal-metal *d-d* orbital hybridization, which is responsible for more ductile behavior, similarly to another system TiTMN [28]. Based on the trends for the $B/G$ ratio and Cauchy pressure, $C_{12}$-$C_{44}$, the promising candidates for improving ductility by alloying include V, Nb, Ta, Mo, and W. Figure. 3c summaries qualitatively the alloying trends on mechanical behavior of CrN, and as such can be used as a guide for tailoring the coating properties.

## 4. Conclusions

A systematic study of the TM-alloying effect to CrN coatings on the lattice parameters, isostructural mixing enthalpies and mechanical properties has been performed using first-principles calculations. The bowing parameters for the lattice parameters of $Cr_{1-x}TM_xN$ (TM=Sc, Y; Ti, Zr, Hf; V, Nb, Ta; Mo, W) in the cubic B1 structure are obtained and the positive deviation from Vegard's linear behavior has been found for $Cr_{1-x}TM_xN$ (TM=Y; Zr, Hf; Nb, Ta) solid solutions, while $Cr_{1-x}TM_xN$ (TM=Sc; Ti; V; Mo, W) solid solutions show a negative bowing. The large lattice mismatch between CrN and YN induces maximum mixing enthalpy of $Cr_{1-x}Y_xN$ and finally results in a strong preference for phase separation, while $Cr_{1-x}Ta_xN$ shows a negative mixing enthalpy in the whole compositional range, indicating that TaN and CrN binaries are miscible. $Cr_{1-x}Sc_xN$ and $Cr_{1-x}V_xN$ have their mixing enthalpy close to zero due to the mutual compensation between electronic effect and the lattice strain. The $B/G$ ratio and Cauchy pressure, $C_{12}$-$C_{44}$, of the ternary $Cr_{0.89}TM_{0.11}N$ solid solutions show a strong dependence on valence electron concentration. V, Nb, Ta, Mo, and W have been shown to be promising candidates for improving ductility of CrN, which is the result of increasing metal-metal $d$-$d$ orbital hybridization.

## Acknowledgements

The financial support by the START Program (Y371) of the Austrian Science Fund (FWF) and the Austrian Federal Ministry of Economy, Family and Youth and the National Foundation for Research, Technology and Development is gratefully acknowledged. The computational results presented have been achieved in part using the Vienna Scientific Cluster (VSC).

Table 1. The bowing parameters, *b*, for each cubic Cr$_{1-x}$TM$_x$N phase together with the optimized lattice constants $a_0$, experimental values, $a_{exp}$, and formation energies, $E_f$ and VEC for different binary cubic TM compounds (TM = Sc, Y, Ti, Zr, Hf, V, Nb, Ta, Cr, Mo and W).

|  | ScN | YN | TiN | ZrN | HfN | VN | NbN | TaN | CrN | MoN | WN |
|---|---|---|---|---|---|---|---|---|---|---|---|
| *b* (Å) | -0.018 | 0.138 | -0.017 | 0.155 | 0.152 | -0.005 | 0.079 | 0.11 |  | -0.062 | -0.024 |
| $a_0$ (Å) | 4.52 | 4.921 | 4.253 | 4.618 | 4.538 | 4.127 | 4.454 | 4.423 | 4.146 | 4.357 | 4.367 |
| $a_{exp}$ (Å) | 4.44[a] | 4.894[a] | 4.241[a] | 4.578[a] | 4.525[a] | 4.139[a] | 4.389[a] | 4.358[a] | 4.135[a] | 4.34[b] | 4.25[b] |
| $E_f$(eV/atom) | -1.923 | -1.698 | -1.727 | -1.686 | -1.758 | -0.983 | -0.874 | -0.854 | -0.441 | 0.025 | 0.329 |
| VEC | 8 | 8 | 9 | 9 | 9 | 10 | 10 | 10 | 11 | 11 | 11 |

[a]Reference 36

[b]Reference 37

**Figure caption**

Fig. 1. (Color online) Calculated equilibrium lattice parameters in comparing with experimental results for $Cr_{1-x}TM_xN$ (TM=Sc, Y, Ti, Zr, Hf, V, Nb, Ta, Mo and W) solid solutions as a function of TMN content. The dotted lines indicate Vegard's linear behavior. Open symbols correspond to the experimental data. The figure is divided into a, b, c, and d, according to the TM groups, IIIB, IVB, VB, and VIB, respectively.

Fig. 2. (Color online) Calculated isostructural mixing enthalpy of $Cr_{1-x}TM_xN$ (TM=Sc, Y, Ti, Zr, Hf, V, Nb, Ta, Mo, and W) solid solutions as a function of TMN content. The figure is divided into a, b, c, and d, according to the TM groups, IIIB, IVB, VB, and VIB, respectively.

Fig. 3. (Color online) (a) The bulk to shear modulus ratio, $B/G$, versus VEC, (b) VEC effect on Cauchy pressure, $C_{12}$-$C_{44}$, in CrN based nitrides, and (c) the alloying-related trends in ductility of $Cr_{1-x}TM_xN$ as functions of TM elements and also TM content. The red solid circle denotes CrN.

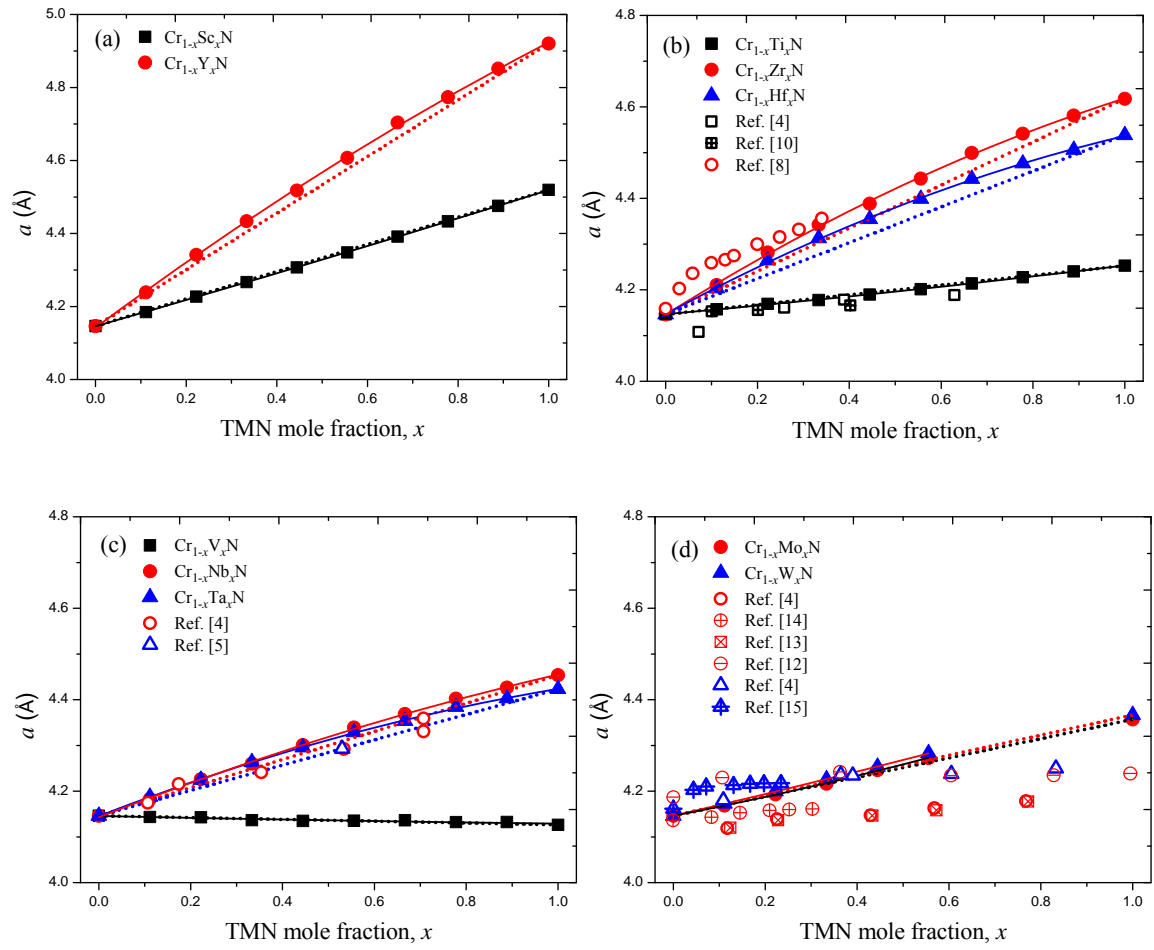

Fig. 1.

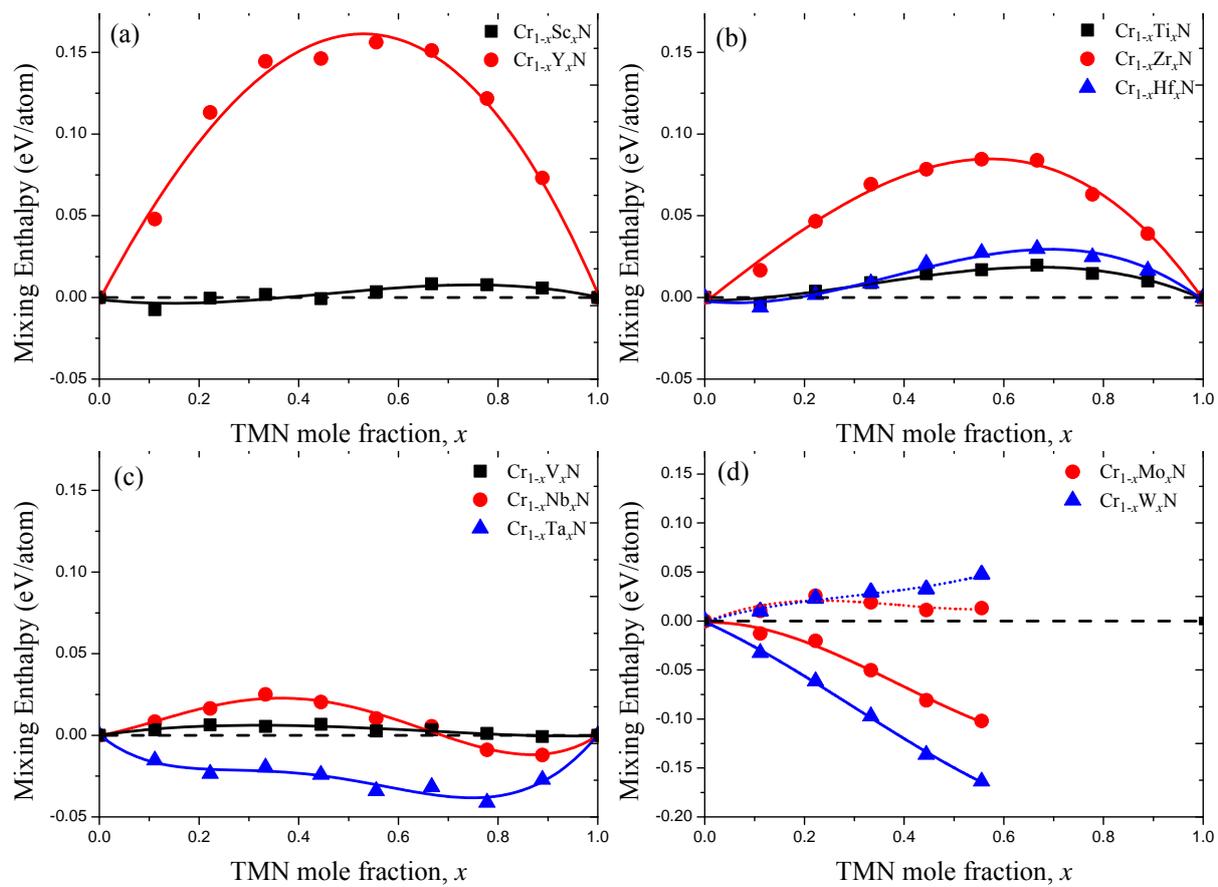

Fig. 2.

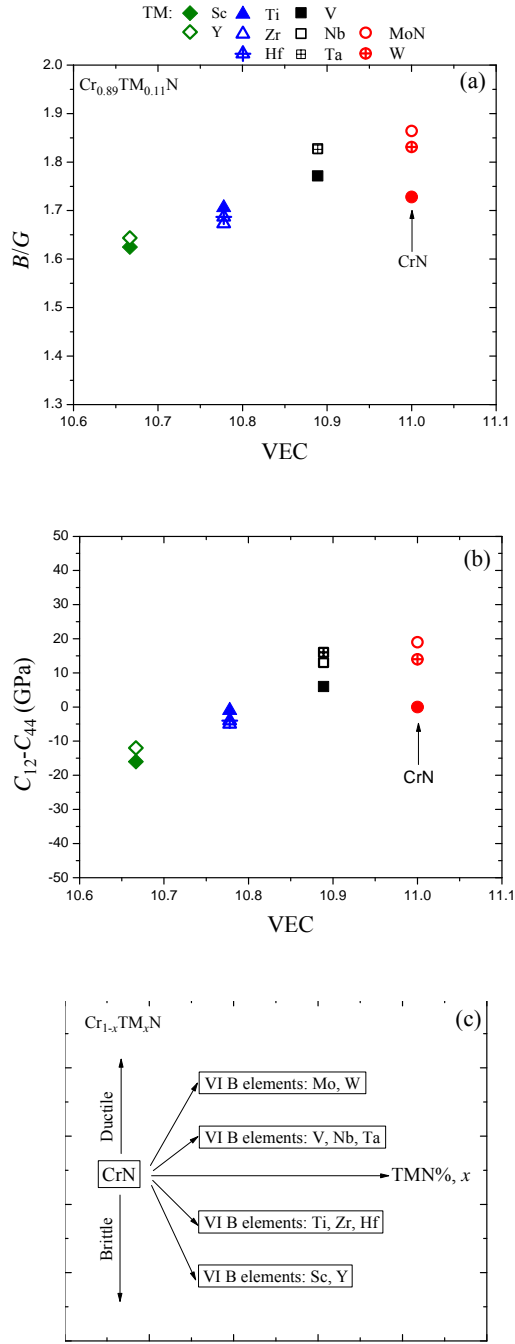

Fig. 3.